\documentclass[a4paper, 12pt]{article}
\usepackage{amsmath, amssymb, graphics, setspace}
\usepackage{cite}
\usepackage{rotating}
\usepackage{longtable}
\newcommand{\mathsym}[1]{{}}
\newcommand{\unicode}[1]{{}}

\catcode`\@=11

\@addtoreset{equation}{section} \catcode`\@=10

\def\d{\mbox{\rm d}}

\def\dddot#1{\mathinner{\buildrel\vbox{\kern5pt\hbox{...}}\over{#1}}}
\def\ddddot#1{\mathinner{\buildrel\vbox{\kern5pt\hbox{....}}\over{#1}}}

\begin{document}

\begin {center}
{\Large Symmetries and reductions of order for certain nonlinear third and second-order differential equations with arbitrary nonlinearity}\\[3 mm]
{\small KM Tamizhmani$ ^ 1$, K Krishnakumar$ ^ 1$ and PGL Leach$ ^ 2$\\[3 mm]
$^ 1 $ Department of Mathematics, Pondicherry University, Kalapet, \\Puducherry 605 014, India\\$^2 $Department of Mathematics, Durban University of Technology, POBox 1334,\\ Durban 4000,~ Republic of South Africa.}
\end{center}
\begin {abstract}
We examine the reductions of the order of certain third- and second-order nonlinear equations with arbitrary nonlinearity through their symmetries and some appropriate transformations. We use the folding transformation which enables one to change from a nonlinearity with an  arbitrary exponent to a nonlinearity with a specific numerical exponent.
\end {abstract}

\section {Introduction}
Nonlinearity plays an important role in various fields of mathematics, physics, biology {\it etc}. The determination of closed-form solution of nonlinear ordinary differential equations is quite difficult in general. Around the turn of twentieth century Painlev\'e and his school classified second-order nonlinear ordinary differential equations based on what is known as the Painlev\'e Property. An ordinary differential equation is said to possess the Painlev\'e Property if all solutions are single-valued around all movable singularities. They arrived at fifty canonical equations of which the solutions of six equations could not be expressed in terms of then known functions. These six equations are the famous Painlev\'e equations \cite{Ince}. There exists another class of equations which are linearizable through transformations of dependent/or independent variable. The famous equation among this class is the Gambier system which is written in the form of coupled Riccati equations in cascade. The Gambier equation can be rewritten as second-order nonlinear ordinary differential equation of the form  \cite{Gambier09a}
\begin{equation}\label{GAM1}
y^{\prime\prime}=\left(I- \frac{I}{n}\right)\dfrac{{y^{\prime}}^2}{y}+\left(a y+b+\frac{c}{y}\right)y^{\prime}- \frac{n a^2}{(n+2)^2}y^3+e y^2+f y+g+ \frac{h}{y},
\end{equation}
where here and below, unless otherwise indicated, the prime denotes differentiation with respect to the independent variable, $x$. For each $n$ one can obtain a condition on $a, b$ and $c$  which guarantees the Painlev\'e Property ({\it ie} absence of movable critical points). There were detailed analyses to determine integrable/linearizable nonlinear ordinary differential equations by Painlev\'e, Gambier, Garnier, Bureau \cite{Painleve97, Painleve00a, Painleve02a, Painleve06a, Garnier12a, Garnier2, Bureau64a} and others through the analysis of their singularity structures. In another development Sophus Lie formulated criteria for linearising certain classes of nonlinear second-order differential equations through the approach of symmetry \cite{Lie74a}. Later Chazy \cite{Chazy11a} gave conditions for linearisation and integration of certain higher-order nonlinear ordinary differential equations. Many authors have been using various methods such as symmetry analysis, the method of the Jacobi Last Multiplier \cite{Nucci02, Nucci04, Nucci05a, Nucci05b, Nucci07, Nucci08, Nucci10}, the Prelle-Singer method \cite{Lakshmanan05, Lakshmanan06a, Lakshmanan06b, Lakshmanan10} {\it etc} to find the solutions/first integrals of nonlinear ordinary differential equations.  Among these methods that of symmetry analysis has been used by researchers effectively \cite{Winternitz02, Winternitz09a, Winternitz09b}. Recently finding the interconnection among these methods has also been a very interesting part of research \cite{Mohana14, Mohana15}.

\strut\hfill

Our main objective of this paper is to investigate the third-order nonlinear ordinary differential equation of Chazy \cite{Chazy11a} through symmetry analysis. From the point of view of symmetry we look for reductions of this equation to well-known equations. We show that one of the reductions of the Chazy equation is an Abel equation of the second kind. Our results motivated us to reexamine the list of new second-order nonlinear ordinary differential equations given in a series of papers \cite{Lakshmanan06a, Lakshmanan10}. We show through the approach of symmetry and the implementation of a series of transformations that all these equations can be transformed to well-known equations which are found in the literature. We compute the Lie point symmetries of these equations by using the Mathematica add-on, Sym \cite {Dimas05a, Dimas06a, Dimas08a, Andriopoulos09a}. One can reduce the order of the given differential equations by using canonical variables which are found from the symmetries of the differential equations. Based on this idea we firstly give the reduction of a third-order ordinary differential equation which was used by Chazy \cite{Chazy11a}. Although he reduced the third-order equation to an Abel equation of the first kind, we reduce it to an Abel equation of the second kind\footnote{The two are kinds of Abel's equations are related by a transformation.}. Although the Abel equation of the second kind is not integrable for arbitrary parameters, there exist particular solutions of them with certain restrictions on the parameters.  These restrictions are already given in \cite{Polyanin03}.

\strut\hfill

Secondly we examine the general form of certain second-order nonlinear equations,
\begin{equation}\label{LAK1}
y^{\prime\prime}+(k_1y^q+k_2)y^{\prime}+k_3y^{2q+1}+k_4y^{q+1}+k_5 y=0
\end{equation}
and
\begin{equation}\label{RLAK1}
y^{\prime\prime}+\lambda\dfrac{{y^{\prime}}^2}{y}+(k_1y+k_2)y^{\prime}+K_3y^3+K_4y^2+K_5 y=0,
\end{equation}
where $q$, $\lambda$, $k_1$, $k_2$, $k_3$, $k_4$, $k_5$, $K_3$, $K_4$ and $K_5$ are arbitrary parameters. For particular values of the parameters in these equations one can obtain many well-known equations of oscillators such as the modified Emden equation, Emden equation {\it etc.}\cite{Lakshmanan05}. Eq.(\ref{LAK1}) and eq.(\ref{RLAK1}) have already been analyzed \cite{Lakshmanan06a, Lakshmanan10} separately through the extended Prelle-Singer method. By this method the authors of \cite{Lakshmanan06a, Lakshmanan10} claimed that the following equations with some restrictions on the parameters are new integrable equations. The first five equations (\ref{LAKN1}--\ref{LAKN5}) are from \cite{Lakshmanan06a}.  The remaining (\ref{LAKN6}--\ref{LAKN10}) are from \cite{Lakshmanan10}.  The equations are
\begin{equation}\label{LAKN1}
y^{\prime\prime}+k_1yy^{\prime}+k_3y^3+k_5 y=0,
\end{equation}
\begin{equation}\label{LAKN2}
y^{\prime\prime}+(k_1y+k_2)y^{\prime}+k_3y^{3}+{k_1k_2\over 3}y^{2}+{2{k_2}^2\over 9} y=0,
\end{equation}
where $k_3={(r-1){k_1}^2 \over 2 r^2}.$
\begin{equation}\label{LAKN3}
y^{\prime\prime}+(k_1y^2+k_2)y^{\prime}+k_3y^{5}+{4(r-1){k_1k_2} \over 3 r^2}y^{3}+{(r-1){k_2}^2 \over r^2} y=0,
\end{equation}
\begin{equation}\label{LAKN4}
y^{\prime\prime}+(k_1y^2+k_2)y^{\prime}+{{k_1}^2\over 16}y^{5}+{{k_1k_2}\over 4}y^{3}+k_5 y=0,
\end{equation}
\begin{equation}\label{LAKN5}
y^{\prime\prime}+(k_1y^2+k_2)y^{\prime}+k_3y^{5}+{{k_1k_2}\over 4}y^{3}+{{3k_2}^2\over 16} y=0,
\end{equation}
where $k_3={(r-1){k_1}^2 \over 3 r^2}$ and $r$ is an arbitrary parameter.
\begin{equation}\label{LAKN6}
y^{\prime\prime}+\lambda\dfrac{{y^{\prime}}^2}{y}+(k_1y+k_2)y^{\prime}+K_3y^3+{k_2K_3(3+2\lambda)\over k_1(1+\lambda)}y^2+{K_3{k_2}^2(2+\lambda)\over {k_1}^2(1+\lambda)} y=0,
\end{equation}
\begin{equation}\label{LAKN7}
y^{\prime\prime}+\lambda\dfrac{{y^{\prime}}^2}{y}+(k_1y+k_2)y^{\prime}+{(1+\lambda){k_1}^2\over (3+2\lambda)^2}y^3+{k_1k_2\over (3+2\lambda)}y^2+K_5 y=0,
\end{equation}
\begin{equation}\label{LAKN9}
y^{\prime\prime}+\lambda\dfrac{{y^{\prime}}^2}{y}+k_2y^{\prime}+K_4y^2+{2(3+2\lambda){k_2}^2\over (5+4\lambda)^2} y=0,
\end{equation}
\begin{equation}\label{LAKN10}
y^{\prime\prime}+\lambda\dfrac{{y^{\prime}}^2}{y}+(k_1y+k_2)y^{\prime}+K_3y^3+{k_1k_2\over (3+2\lambda)}y^2+{(2+\lambda){k_2}^2\over (3+2\lambda)^2} y=0.
\end{equation}

 Our investigation reveals that all the above new integrable equations can be identified with well-known equations, such as the Riccati equation, Bernoulli equation, Abel equations of the first and second kinds and the Emden-Fowler equation, by taking a series of transformations that are already given in \cite{Polyanin03}. As the determination of these solutions and conditions for linearisation of equations of these types are well-known, we can show that the equations listed above are not essentially new equations but variations on the types mentioned above. For both third- and second-order equations we use appropriate folding transformations\footnote{Okamoto {\it et al} \cite{Okamoto} introduced the folding transformation.  This is a class of transformations of algebraic rather than rational form. This type of transformation plays a crucial role in many of continuous and discrete integrable systems \cite{Grammaticos, Partha15} } which takes arbitrary exponents to a specific numerical nonlinearity. Otherwise finding the symmetries of these equations with arbitrary nonlinearity is difficult.

\strut\hfill

\section{Symmetries and reductions of the third-order equation}
We consider equation of the form \cite{Gambier09a}
\begin{equation}\label{Third3}
-y^{\prime\prime\prime}+A y^q y^{\prime\prime}+B y^{q-1}{y^{\prime}}^2+F y^{2q}y^{\prime}+G y^{3q+1}=0.
\end{equation}
When one applies the folding transformation, $v(x)=y^{-q},$ one gets the form
\begin{eqnarray}\label{Third}
v^2 v^{(3)}-3\left(1+\frac{1 }{q}\right)v v^{\prime}v^{\prime\prime}-A v v^{\prime\prime}+\left(2 +\frac{1}{q^2}+\frac{3 }{q}\right){v^{\prime}}^3&&\nonumber\\ +\left(A +\frac{A }{q}+\frac{B}{q}\right){v^{\prime}}^2-F v^{\prime}+G q=0.&&
\end{eqnarray}
Now the arbitrary nonlinearity of (\ref{Third3}) has been removed by the above folding transformation. The symmetries of (\ref{Third}) are
\begin{eqnarray}
&&\Gamma_1=\partial_x\\
&&\Gamma_2=x\partial_x+v \partial_v.
\end{eqnarray}
According to the symmetry $\Gamma_1$ the canonical variables are $(r=v,\ R = v').$ By using these canonical variables one can find the transformation $$v=r,\ v^{\prime}=R(r)$$ to reduce the third-order nonlinear ordinary differential equation, (\ref{Third}), to a second-order nonautonomous nonlinear ordinary differential equation given by
\begin{eqnarray}\label{Second1}
&&q^2 r^2( R
{R^{\prime}}^2+ R^2
R^{\prime\prime})-3 q r(1+q) R^2 R^{\prime}-A q^2 r R R^{\prime}\nonumber\\&&+(1+3 q +2 q^2 )R^3+(A q +B q +A q^2 )R^2-F q^2 R+G q^3=0,
\end{eqnarray}
where here $^{\prime}$ represents differentiation with respect to $r.$  The symmetry of (\ref{Second1}) is given by
\begin{equation}
\Gamma_1=r\partial_r,
\end{equation}
and corresponding canonical variables are $(z=\log r,\ Z=R).$  Thereby we obtain the transformation $$R=Z,\ r R^{\prime}=Z^{\prime}$$ and the second-order autonomous ordinary differential equation
\begin{eqnarray}\label{Second2}
q Z^2 \left(q \left(Z^{\prime \prime}-Z^{\prime}\right)-3 (1+q) Z^{\prime}+A q+A+B\right)+\left(1+3 q+2 q^2\right) Z^3&&\nonumber\\ -q^2 Z \left(F+A Z^{\prime}-{Z^{\prime}}^2\right)+G q^3=0,&&
\end{eqnarray}
where now $^{\prime}$ represents differentiation with respect to $z.$ The symmetry of (\ref{Second2}) is given by
\begin{equation}
\Gamma_1=\partial_z.
\end{equation}
The symmetry leads to the canonical variables $(h=Z,\ z=T)$ and the transformations $$Z=h,\ Z^{\prime}=H$$ which reduce equation (\ref{Second2}) to an Abel's equation of the second kind,
\begin{equation}\label{AbelSecondKind}
H H^{\prime}+\frac{H^2}{h}-\frac{A H}{h}-\left(4+\frac{3}{q}\right)H+\frac{G q}{h^2}-\frac{F}{h}+2 h+\frac{h}{q^2}+\frac{A}{q}+\frac{B}{q}+\frac{3 h}{q}+A=0,
\end{equation}
where $^{\prime}$ represents differentiation with respect to $h.$ The Abel's equation of the second kind (\ref{AbelSecondKind}) is not integrable for arbitrary parameters. However, we give particular solutions of this equation with certain possible restrictions among the parameters. The details can be found in \cite{Polyanin03}. For the sake of completeness we reproduce the results given in \cite{Polyanin03} for the various cases of (\ref{AbelSecondKind}) listed below. Here we excluded the trivial linear cases.
\bigskip
\strut\hfill
\\[3mm]
Case 1:\\ \smallskip

$A=\text{arbitrary},$ $B=-\dfrac{A}{4},$ $F = \text{arbitrary},$ $G = 0,$ $q = -\dfrac{3}{4}$
\begin{equation}
 H H^{\prime}+ \dfrac{H^2}{h} - \dfrac{A H}{h} -\dfrac{2h}{9} -\dfrac{F}{h} =0.
\end{equation}
The transformation $H=A \dfrac{w(h)}{h}$ [see \cite{Polyanin03}, 1.3.4-2, equation 6] reduces this to
\begin{eqnarray}\label{Emden1}
&&w w^{\prime}-w - \dfrac{2h^3}{9A^2} -\dfrac{F h}{A^2}=0.
\end{eqnarray}
One has a form of the Emden-Fowler equation
\begin{eqnarray}\label{Emden2}
&&\phi^{\prime\prime}=C \xi^n \phi^3
\end{eqnarray}
from (\ref{Emden1}) by the transformation [see \cite{Polyanin03}, 1.3.1-2 equation 71]
\[
h =\phi \xi^{1 + {n \over 2}}, w = \xi^{1 + {n \over 2}} \dfrac{(2 + n) \phi + 2 \xi \phi^{\prime}}{2 (3 + n)},
\]
where
\[
\dfrac{F}{A^2}= -\dfrac{(2 + n) (4 + n)}{4 (3 + n)^2} \quad \Leftrightarrow \quad n =
 \pm  \left(\dfrac{A}{\sqrt{A^2 + 4 F}}-3\right),
\]
the prime denotes differentiation with respect to $\xi$ and $C=-{2 (3 + n)^2/(9 A^2)}$. \smallskip
\strut\hfill
\\[3mm]
 Subcase: $n=-6$\\
 \smallskip

 The solution of (\ref{Emden2}) in parametric form is
 \begin{equation}
 \xi=a I_1^2\left[\int(1\pm\tau^4)^{-1\over 2}d\tau+I_2\right]^{-1},\quad \phi=b I_1^4\tau \left[\int(1\pm\tau^4)^{-1\over 2}d\tau+I_2\right]^{-1},
 \end{equation}
 where $C=\pm 2 a^4b^{-2}.$\\

 \strut\hfill

This result is a consequence of the property that for $n = -6$ (\ref{Emden2}) has the two Lie point symmetries,
\[
\Gamma_1 = \xi\partial_{\xi} + 2\phi\partial_{\phi} \quad \mbox{\rm and} \quad \Gamma_2 = \xi^2\partial_{\xi} + \xi\phi\partial_{\phi},
\]
whereas in general there is just the single symmetry $\Gamma = 2\xi\partial_{\xi} - (2+n)\phi\partial_{\phi}$.

\strut\hfill
\\[3mm]
Case 2:\\ \smallskip

$A=-2 B,$ $ B=\text{arbitrary},$ $F = \text{arbitrary},$ $ G = 0,$  $q = -\dfrac{3}{2}$
\begin{eqnarray}
&& H H^{\prime}+ \dfrac{H^2}{h} - \dfrac{2B H}{h}-2H +\dfrac{4h}{9} -\dfrac{F}{h}-\dfrac{4B}{3} =0.
\end{eqnarray}
The transformation $H=w(h)+\dfrac{2}{3}h$ [see \cite{Polyanin03}, 1.3.4-2, equation 7] reduces this equation to
\begin{eqnarray}
&&(F - 2 B w - w^2) \dfrac{dh}{dw} = w h + \dfrac{2}{3} h^2
\end{eqnarray}
with the solution
\begin{eqnarray}
   &&h=-\dfrac{3 \exp{\left(L\tan^{-1}M\right)} F}{
 4 B \exp{\left(L\tan^{-1}M\right)} +
  2 \exp{\left(L\tan^{-1}M\right)} w -
  3 F N I},
\end{eqnarray}
where $I$ is a constant of integration,
\[
L=\dfrac{B}{\sqrt{-B^2 - F}}, \quad M=\dfrac{(B + w)}{\sqrt{-B^2 - F}} \quad \mbox{\rm and} \quad N=\sqrt{-F + 2 B w + w^2}.
\]
\strut\hfill
\\[3mm]
Case 3:\\ \smallskip

$A=\dfrac{5}{2},$ $ B=\text{arbitrary},$ $F = -\dfrac{3}{2},$ $G = 0,$ $q = 1$
\begin{eqnarray}
&& H H^{\prime}+ \dfrac{H^2}{h} - \dfrac{5 H}{2h}-7H +6 h +\dfrac{3}{2h}+B+5 =0.
\end{eqnarray}
The transformation $h=\xi^2$ and $H=\xi w(\xi)+2\xi^2+1$ [see \cite{Polyanin03}, 1.3.4-2, equation 8] reduces this equation to
\begin{eqnarray}
&&(2 - 2 B - 3 w^2) \dfrac{d\xi}{dw} = 1+w \xi +2 \xi^2
\end{eqnarray}
with solution
\begin{eqnarray}
&&\xi=\left(-24 \sqrt{-1 + B}
     w I ~~P\left[l,m, n\right] + \sqrt{11 - 10 B - B^2}) I ~~P\left[7l,m, n\right]\right.\nonumber\\ && \left. -
   24 \sqrt{-1 + B}w ~~Q\left[l,m, n\right] -7 \sqrt{6} \sqrt{-(-1 + B)^2} \left(1+ ~~Q\left[7l, m, n\right] \right)\right.\nonumber\\ && \left.+ \sqrt{6} \sqrt{11 - 10 B - B^2}~~ Q\left[7l, m, n\right]\right)/\nonumber\\&&\left(12 \sqrt{-1 + B}I ~~P\left[l,m, n\right]+ 12 \sqrt{-1 + B}~~
      Q\left[l, m, n\right]\right),
\end{eqnarray}
where $P$ and $Q$ are Legendre functions, $I=$ is a constant of integration,
\[
l=\dfrac{1}{6}, \quad m=\dfrac{1}{6}  \sqrt{\dfrac{(11 + B)}{(-1 + B)}} \quad \mbox{\rm and} \quad n=-\dfrac{\sqrt{3} w}{\sqrt{2 (1 - B)}}.
\]

\strut\hfill
\\[3mm]
Case 4:\\ \smallskip

$A=\dfrac{5}{2},$ $B=\text{arbitrary},$ $F = -\dfrac{3}{2},$ $G = 0,$  $q = -\dfrac{5}{2}$
\begin{eqnarray}
&& H H^{\prime}+ \dfrac{H^2}{h} - \dfrac{5 H}{2h}-\dfrac{14H}{5} +\dfrac{24h}{25} +\dfrac{3}{2h}-\dfrac{2B}{5}+\dfrac{3}{2} =0.
\end{eqnarray}
The transformation $h=\xi^2$ and $H=\xi w(\xi)+\dfrac{4}{5}\xi^2+1$ [see \cite{Polyanin03}, 1.3.4-2, equation 8] reduces this to
\begin{eqnarray}
&&(9 + 4 B - 15 w^2) \dfrac{d\xi}{d w} = 5+5w \xi +4 \xi^2
\end{eqnarray}
with the solution
\begin{eqnarray}
\xi&=& -\left(120 w I ~~P\left[i,j, k\right]+ 120 w~~ Q\left[i,j, k\right]\right.\nonumber \\
       && \left.+ \sqrt{15} (\sqrt{-39 + 4 B} - 7 \sqrt{9 + 4 B}) I~~ P\left[i,j, k\right]\right. \nonumber \\
       &&\left. - 7 \sqrt{15} \sqrt{9 + 4 B}~~Q\left[7i,j, k\right] + \sqrt{-585 + 60 B}~~
        Q\left[7i,j, k\right]\right)/  \nonumber\\
       && \left(24 I ~~P\left[i, j, k\right]+24 ~~Q\left[i,j, k\right]\right),
\end{eqnarray}
where $P$ and $Q$ are Legendre functions, $I=$ is a constant of integration, $i=\dfrac{1}{6},$ $j=\dfrac{1}{6} \sqrt{\dfrac{(-39 + 4 B)}{(9 + 4 B)}}$  and $k=\dfrac{\sqrt{15} w}{\sqrt{9 + 4 B}}.$
\bigskip
\strut\hfill
\\[3mm]
Case 5:\\ \smallskip

$A=0,$ $B=0,$ $F = \text{arbitrary},$ $G = 0,$ $ q = -\dfrac{3}{4}$
\begin{eqnarray}
&& H H^{\prime}+ \dfrac{H^2}{h} - \dfrac{2h}{9}  -\dfrac{F}{h} =0,
\end{eqnarray}
the solution of which is given by
\begin{eqnarray}
&&H= \pm \dfrac{\sqrt{9F h^2+h^4+9I}}{3h}.
\end{eqnarray}
\bigskip
\\[3mm]
Case 6:\\ \smallskip

$A = -2 B,$ $B=\text{arbitrary},$ $F = 0,$ $G = 0,$  $q = -\dfrac{1}{2}$
\begin{eqnarray}
&&  H^{\prime}+ \dfrac{H}{h} + \dfrac{2B}{h}  +2 =0
\end{eqnarray}
with solution given by [see \cite{Polyanin03}, 1.3.4-2, equation 34]
\begin{eqnarray}
&&H=  -2 B - h + \dfrac{I}{h}.
\end{eqnarray}
\strut\hfill
\\[3mm]
Case 7:\\ \smallskip

$A=\text{arbitrary},$ $B=0,$ $F = 0,$ $G = 0,$  $q = -1$
\begin{eqnarray}
&&  H^{\prime}+ \dfrac{H}{h} - \dfrac{A}{h}  -1 =0.
\end{eqnarray}
This equation has the solution [see \cite{Polyanin03}, 1.3.4-2, equation 34]
\begin{eqnarray}
&&H=  A +\dfrac{h}{2}+ \dfrac{I}{h}.
\end{eqnarray}
\strut\hfill

\section {Symmetries and reductions of the second-order equation}
We consider the equation (\ref{LAK1}),
\begin{equation}\nonumber
y^{\prime\prime}+(k_1y^q+k_2)y^{\prime}+k_3y^{2q+1}+k_4y^{q+1}+k_5 y=0.
\end{equation}
When we apply the folding transformation, $w=y^q,$ we obtain the following form
\begin{equation}\label{14}
w^{\prime\prime}+\left({1 \over q}-1\right)\dfrac{{w^{\prime}}^2}{w}+(k_1 w+k_2)w^{\prime}+ q k_3w^3+q k_4w^2+ q k_5 w=0.
\end{equation}
\textbf{Remark:} \\ When $({1\over q}-1)=\lambda,$ $ { k_3\over 1+\lambda}=K_3,$ $ { k_4\over 1+\lambda}=K_4$ and $ { k_5\over 1+\lambda}=K_5$, the equation above, (\ref{14}), is equation (\ref{RLAK1}). Therefore there is no need to make a separate analysis.\\ \bigskip
\strut\hfill

Now we apply the symmetry method to find the further transformations which reduce the order of equation (\ref{14}). The determining equations of (\ref{14}) have four arbitrary functions and conditions over them. These functions do exist in the coefficient functions, $\xi$ and $\eta$, according to
\begin{eqnarray}
\xi(x,w)&&= a(x)+q w^{1\over q}b(x)\\
\eta(x,w)&&={w^{1-{1\over q}}\over (1 + q) (2 + q)}\left[\left(-q^2 w^{2\over q}
    b(x) (2 w k_1 + (2 + 3 q + q^2) k_2)\right.\right. \nonumber\\ &&\left.\left.+ (2 + 3 q + q^2) (w^{1\over q} d(x) + c(x) +
     q^2 w^{2\over q} b^{\prime}(x))\right)\right],
\end{eqnarray}
where $a(x),b(x),c(x)$ and $d(x)$ are arbitrary functions.  According to the conditions on the arbitrary functions, there do occur a few cases\\[3mm]
Case 1:\\ \smallskip

$a(x) = A_0,$ $b(x)=0,$ $c(x)=0$ and $d(x)=0$\\
 \smallskip

 The corresponding symmetry is
$$\Gamma_1=\partial_x$$
\smallskip
\\[3mm]
Case 2:\\ \smallskip

\[
a(x) = A_1 -   {(2 + q) A_0 \over q k_2}\exp\left[{q k_2 x\over 2 + q}\right],\quad  b(x)=0,\ c(x)=0 \quad \mbox{\rm and} \quad d(x)=A_0 \exp\left[{q k_2 x\over 2 + q}\right]
\]
with the conditions $k_4= {k_1 k_2/(2 + q)}$ and $k_5= {(1 + q){k_2}^2/(2 + q)^2}.$\\
The corresponding symmetries are
\begin{eqnarray}
&&\Gamma_1=\partial_x\\
&&\Gamma_2= -\exp\left[{q k_2 x\over 2 + q}\right]\left({(2 + q) \over q k_2}\partial_x+w\partial_w\right).
\end{eqnarray}
\\[3mm]
Case 3:\\ \smallskip

$k_1=0,$ $k_3=0$ and $k_4=0$.\\
The corresponding symmetries are
\begin{eqnarray*}
&&\Gamma_1=w\partial_w,\\
&&\Gamma_2=\partial_x - \dfrac{1}{2} k_2  q w \partial_w,\\
&&\Gamma_3=\exp\left[-\dfrac{(k_2 + N)x}{2}\right] w^{\left({-1 + q\over q}\right)}\partial_w,\\
&&\Gamma_4=\exp\left[\dfrac{(-k_2 + N)x}{2}\right] w^{\left({-1 + q\over q}\right)}\partial_w,\\
&&\Gamma_5=\exp[N x]\dfrac{(2 \partial_x -   k_2 q w \partial_w + q  N w \partial_w)}{2N},\\
&&\Gamma_6=\exp[-N x]\dfrac{(-2 \partial_x +   k_2 q w \partial_w + q  N w \partial_w)}{2N},\\
&&\Gamma_7=\dfrac{q w^{1/q}}{2}\exp\left[\dfrac{(k_2 + N)x}{2}\right](2 \partial_x -   k_2 q w \partial_w + q  N w \partial_w) \quad \mbox{\rm and}\\
&&\Gamma_8=-\dfrac{q w^{1/q}}{2}\exp\left[\dfrac{(k_2 - N)x}{2}\right] (-2 \partial_x +   k_2 q w \partial_w + q  N w \partial_w),
\end{eqnarray*}
where $N=\sqrt{k_2^2-4k_5}.$
 When we treat Case 1, we arrive at the conditions $k_4= {k_1 k_2/(2 + q)}$ and $k_5= {(1 + q){k_2}^2/(2 + q)^2}$ and the last case contains the conditions $k_1=0, k_3=0$ and $k_4=0$ which means that the given second-order equation is a linear equation. Hence only one reduction is sufficient for all cases by using $\Gamma_1=\partial_x.$

\strut\hfill

According to the symmetry $\Gamma_1$ the canonical variables are $(r=w,\ s=x).$ By using these canonical variables one can find the transformation $$w=r,\ w^{\prime}=H(r)$$ to reduce the second-order nonlinear ordinary differential equation, (\ref{14}), to a first-order nonautonomous nonlinear ordinary differential equation with the form of an Abel's equation of the second kind. It is
\begin{equation}\label{15}
HH^{\prime}+\left({1 \over q}-1\right)\dfrac{H^2}{r}+(k_1 r+k_2)H+ q k_3r^3+q k_4r^2+ q k_5 r=0,
\end{equation}
where $^{\prime}$ represents differentiation with respect to $r.$
The transformation $H(r)=R(r)r^{1-{1/q}}$ changes the Abel equation of the second kind into one of the same kind with new dependent variable $R(r)$. It has the form
\begin{equation}\label{UABEL}
RR^{\prime}+(k_1 r+k_2)r^{{1 \over q}-1}R+ (q k_3r^2+q k_4r+ q k_5 )r^{{2\over q}-1}=0.
\end{equation}

\subsection{Reduction to an Abel equation}
If one takes $k_2=0,$ $k_4=0$  and $q=1,$ then $ (\ref{UABEL})$ reduces to an equation of the form
\begin{equation}\label{SABE1L}
RR^{\prime}+k_1 r R+ ( k_3r^2+  k_5 )r=0.
\end{equation}
We now make the transformation $z=-\mbox{$\frac{1}{2}$}k_1 r^2$ [see \cite{Polyanin03}, 0.1.6-2]. Eq.$(\ref{SABE1L})$ reduces to a solvable Abel equation of the form [see \cite{Polyanin03}, 1.3.1-2, equation 2]
\begin{equation}\label{FIRABEL1}
R\dfrac{dR}{dz}-R={k_5\over k_1}- {2 k_3\over{ k_1}^2}z
\end{equation}
with no constraints upon the values of $k_1,$ $k_3$ and $ k_5$ \footnote{As these parameter occur in ratios, any two are independent.}. The solution in parametric form is
\begin{equation}\label{QUAD1}
z=c ~\exp\left[-\int\displaystyle{{\tau\over\tau^2-\tau+ {2 k_3\over{ k_1}^2}}}\d\tau+{k_1k_5\over2 k_3}\right],
\end{equation}
\begin{equation}\label{QUAD2}
R=c~\tau ~\exp\left[-\int\displaystyle{{\tau\over\tau^2-\tau+ {2 k_3\over{ k_1}^2}}}\d\tau\right].
\end{equation}
When one uses (\ref{QUAD1}) and (\ref{QUAD2}), one can get the solution of $(\ref{LAKN1})$ in the list of the new integrable cases.

\strut\hfill

\subsection{Reduction to a linear first-order equation}
If we make the transformation
\[
R=r^{{1 \over q}+1}\left(t-{qk_2\over (2+q)r}\right),
\]
 $(\ref{UABEL})$ reduces to a solvable linear first-order equation of the form [see \cite{Polyanin03}, 1.3.3-2, equation 13 with $a=0, n=1+{1/q},c=-k_1,\displaystyle{b=-{qk_2/(q+2)}},d=qk_3$]
\begin{equation}\label{FIR1}
\left(\left({1\over q}+1\right)t^2+k_1 t+qk_3\right)\dfrac{dr}{dt}={qk_2\over(2+q)}-t r
\end{equation}
with conditions on the parameters
\[
k_3 - \mbox{\rm arbitrary},\quad k_4=\dfrac{k_1 k_2}{2+q} \quad \mbox{\rm and} \quad k_5=\dfrac{(1+q){k_2}^2}{(2+q)^2}.
\]
The solution of $(\ref{FIR1})$ is [see \cite{Polyanin03}, 0.1.2-5]
\begin{equation}\label{SOL2}
r=\exp\left(-\int\dfrac{t \d t}{f(t)}\right)\left[\int\left(\exp\int\dfrac{t \d t}{f(t)}\right)\left(qk_2\over(2+q)f(t)\right)\d t+C\right],
\end{equation}
where $f(t)=({1+1/q}+1)t^2+k_1 t+qk_3.$ We now show that, given the results presented above, one can precisely obtain the results given in \cite{Lakshmanan06a, Lakshmanan10} for various values of $q.$

\strut\hfill

For example take $q=1.$ Then $(\ref{SOL2})$ gives the solution of $(\ref{LAKN2})$ with an arbitrary value of $k_3,$   $k_4=\dfrac{k_1 k_2}{3}$ and $k_5=\dfrac{2{k_2}^2}{9}$ \cite{Lakshmanan06a}.  For $q=2$  $(\ref{SOL2})$ gives the solution of $(\ref{LAKN5})$ with an arbitrary value of $k_3$, $\displaystyle{k_4=\dfrac{k_1 k_2}{4}}$ and $\displaystyle{k_5=\dfrac{3{k_2}^2}{19}}$ \cite{Lakshmanan06a}. If $\displaystyle{q={1\over 1+\lambda}}$,  then $(\ref{SOL2})$ gives the solution of $(\ref{LAKN10})$ with arbitrary value of $K_3$, $\displaystyle{K_4={k_1k_2\over (3+2\lambda)}}$ and $\displaystyle{K_5={(2+\lambda){k_2}^2\over (3+2\lambda)^2}}$ \cite{Lakshmanan10}.

\strut\hfill

\subsection{Reduction to a solvable Abel equation}
If one takes the transformation
\[
\xi=-k_1r^{{1/q}}\left(\dfrac{q r}{1+q}+\dfrac{qk_2}{k_1}\right),
\]
eq.$(\ref{UABEL})$ is reduced to a solvable Abel's equation of the form [see \cite{Polyanin03}, 1.3.3-2, equation 9 with $a=-k_1, \displaystyle{n={1/q}}, \displaystyle{b=-{qk_2/k_1}}$]
\begin{equation}\label{SABEL1}
R\dfrac{dR}{d\xi}=R-\dfrac{(1+q)k_3}{{k_1}^2}\xi
\end{equation}
with conditions $k_3$--arbitrary, $k_4=\dfrac{(2+q)k_2 k_3}{k_1}$ and $k_5=\dfrac{(1+q){k_2}^2k_3}{{k_1}^2}$.
The solution of $(\ref{SABEL1})$ in parametric form is [see \cite{Polyanin03}, 1.3.1-2, equation 2]
\begin{equation}\label{QUAD3}
\xi=C ~\exp\left(-\int\displaystyle{\dfrac{\tau \d\tau}{\tau^2-\tau+{(1+q)k_3 \over {k_1}^2}}}\right),
\end{equation}
\begin{equation}\label{SOL3}
R=C \tau ~\exp\left(-\int\displaystyle{\dfrac{\tau \d\tau}{\tau^2-\tau+{(1+q)k_3 \over {k_1}^2}}}\right).
\end{equation}
\\ \bigskip

Take $q=2$ and $\displaystyle{k_3={(r-1){k_1}^2/(3 r^2)}}.$  Then $(\ref{SOL3})$ gives the solution of $(\ref{LAKN3})$ with the conditions  $\displaystyle{k_4={4(r-1){k_1k_2} \over 3 r^2}}$ and $\displaystyle{k_5={(r-1){k_2}^2 \over r^2}}$ \cite{Lakshmanan06a}.\\
Take $\displaystyle{q={1\over 1+\lambda}}.$ Then $(\ref{SOL3})$ gives the solution of $(\ref{LAKN6})$ with arbitrary value of $K_3$, $\displaystyle{K_4={k_2K_3(3+2\lambda)\over k_1(1+\lambda)}}$ and $\displaystyle{K_5={K_3{k_2}^2(2+\lambda)\over {k_1}^2(1+\lambda)}}$ \cite{Lakshmanan10}. However, we have obtained solutions (\ref{QUAD3}) and (\ref{SOL3}) for arbitrary $k_3.$

\strut\hfill

\subsection{Reduction to a Bernoulli equation}
If one uses the transformation
\[
R=r^{{1 \over q}}\left(\phi-{q k_1 \over 2+q}\right),
\]
eq.$(\ref{UABEL})$ is reduced to a Bernoulli equation of the form [see \cite{Polyanin03}, 1.3.3-2, equation 10 with $\displaystyle{a=-{qk_1\over(2+q)}}, n={1\over q},b=-k_2$],
\begin{equation}\label{BER1}
\left({\phi^2 \over q}+k_2 \phi+q k_5\right)\dfrac{dr}{d\phi}=-\phi r+{q k_1 \over 2+q}r^2
\end{equation}
with conditions $k_3=\dfrac{{k_1}^2}{(2+q)^2}$, $k_4=\dfrac{k_1 k_2}{(2+q)}$ and $k_5$--arbitrary.
The solution of $(\ref{BER1})$ is [see \cite{Polyanin03}, 0.1.2-6]
\begin{equation}\label{SOL4}
r^{-1}=\exp\left(-\int\dfrac{\phi \d\phi}{g(\phi)}\right)\left[C-\int \exp\left(\int\dfrac{\phi \d\phi}{g(\phi)}\right){q k_1 \over 2+q}\d\phi\right],
\end{equation}
where $g(\phi)={\phi^2 \over q}+k_2 \phi+q k_5.$
\\ \bigskip
We recover the conditions given in \cite{Lakshmanan06a} by taking $q=2$ from (\ref{UABEL}).  Then $(\ref{SOL4})$ gives the solution of $(\ref{LAKN4})$ with an arbitrary value of $k_5$ and the conditions
 $k_3=\dfrac{{k_1}^2}{16}$ and $k_4=\dfrac{k_1 k_2}{4}$ \cite{Lakshmanan06a}. Take $q=\displaystyle{{1\over 1+\lambda}}.$ Then $(\ref{SOL4})$ gives the solution of $(\ref{LAKN7})$ with an arbitrary value of $K_5,$ $\displaystyle{K_3={(1+\lambda){k_1}^2\over (3+2\lambda)^2}}$ and $\displaystyle{K_4={k_1k_2\over (3+2\lambda)}}$ \cite{Lakshmanan10}.

\strut\hfill

\subsection{Reduction to a Riccati equation}
If one takes the transformation
\[
\chi=\sqrt{r}~ \text{and} ~ R=r^{2+q \over 2q}\left(t-{2q k_1\sqrt{r}\over 4+3 q}-{2 q k_2 \over (4+q)\sqrt{r}}\right),
\]
eq.$(\ref{UABEL})$ is reduced to a Riccati equation of the form [see \cite{Polyanin03}, 1.3.3-2, equation 11],
\begin{equation}\label{RIC1}
\left[\left({2+q\over 2q}\right)t^2+c_0\right]\dfrac{d\chi}{dt}={qk_1\over 4+3q}\chi^2-{t\over 2}\chi+{qk_2\over 4+q},
\end{equation}
where
\[
c_0=qk_4-{4q(2+q)k_1k_2 \over (4+q)(4+3q)}
\]
with conditions
\[
k_3=\dfrac{2(2+q){k_1}^2}{(4+3q)^2},\quad k_4 \mbox{\rm -arbitrary and} \quad k_5=\dfrac{2(2+q){k_2}^2}{(4+q)^2}.
\]
The Cole-Hopf transformation,
\[
\chi= \displaystyle{{-\left({2+q\over 2q}t^2+c_0\right)\mu^{\prime}\over {qk_1\over 4+3q}\mu}},
\]
converts the nonlinear Riccati equation $(\ref{RIC1})$ into a linear second-order equation of the form:
\begin{equation}\label{LS1}
\left({2+q \over 2q}t^2+c_0\right)^2\mu^{\prime\prime}_{tt}+\left({2+q \over q}+{1\over 2}\right)t\left({2+q \over 2q}t^2+c_0\right)\mu^{\prime}_t+{q^2k_1k_2 \over (4+q)(4+3q)}\mu=0.
\end{equation}
Take $\displaystyle{q={1\over 1+\lambda}}$ and $k_1=0.$ Then the solution of $(\ref{LS1})$ gives the solution of $(\ref{LAKN9})$ with arbitrary value of $K_4$ and $\displaystyle{K_5={2(3+2\lambda){k_2}^2\over (5+4\lambda)^2}}$ \cite{Lakshmanan10}.

\strut\hfill

\section {Conclusion}

We have investigated the linearisation of two classes of nonlinear equations, one class being of the third order and the other of the second order.  These equations contain parameters.  By the performance of a symmetry analysis we have been able to identify those values of the parameters which permit the existence of Lie point symmetries and consequently reduction of order.  In particular we were able to identify third-order equations with two Lie point symmetries and this led to reduction to first-order equations generally with the form of an Abel's equation of the second kind.  Integrable cases of this class of equations are provided in the compendium of Polyanin and Zeitev \cite{Polyanin03}.

\strut\hfill

In the case of the second-order equations we were able to produce a uniform examination of equations reported in earlier literature \cite{Lakshmanan05, Lakshmanan06a, Lakshmanan06b, Lakshmanan10} as being new additions to the pantheon of integrable second-order equations.  Whilst it is true that these equations were of new appearance, we showed that they could be transformed into equations of known integrability.  However, there is no denying that they do provide new expressions for integrable equations.

\section*{Acknowledgements}
KMT would like to thank Professor Tudor Ratiu, CIB, EPFL for the invitation and hospitality where the work was initiated. KK thanks the University Grants Commission for providing a UGC-Basic Scientific Research Fellowship to perform this research work.  PGLL thanks Professor KM Tamizhmani and the Department of Mathematics, Pondicherry University, for the invitation and provision of facilities while this work was undertaken.

\end{document}